\documentclass[preprint2]{aastex}
\usepackage[dvips]{epsfig}

\shorttitle{Correlation Integral Characterization}
\shortauthors{Barnes}

\begin{document}
\title{Using Correlation Integrals to Characterize 3D Stellar Orbits} 
\author{Eric I. Barnes}
\affil{Department of Physics \& Astronomy, Louisiana State University, Baton Rouge, 
LA 70803}
\email{barnes@physics.lsu.edu}

\begin{abstract}
\indent
In an effort to more fully understand the variety of stellar distribution 
functions that can be used to construct models of realistic galaxies, 
the correlation integral method for orbit characterization that previously 
has been introduced by \citet{gnp83} and \citet{cns84} is examined in considerable
detail.  The broad utility of the method is validated and demonstrated by applying 
it to orbits in a number of different, previously studied test cases (1D, 2D, and 3D; 
nonrotating and rotating).  At the same time, the correlation
integral method is compared and contrasted with other more traditional 
characterization tools such as  Lyapunov exponents and surfaces of section.
The method is then extended to orbits in a previously unexamined rotating,
3D bar potential.  The correlation integral method is found to be a
simple and reliable way to quantitatively categorize orbits in virtually
any potential.  It is recommended that it be broadly adopted as a tool for 
characterizing the properties of orbits and, by extension, stellar 
distribution functions, in all Hamiltonian dynamical systems.
\end{abstract}

\section{Introduction}
\subsection{Background}\label{backg}
\indent
The study of orbits in three-dimensional (3D) gravitational potentials is of broad astrophysical 
interest.  Whether the orbits under investigation are those of asteroids in the solar system (e.g.,
Pilat-Lohinger {\it et al.} 1999), stars in a globular cluster or a galaxy (e.g., Carpintero 
{\it et al.} 1999), or galaxies in a cosmological simulation (e.g., Colpi {\it et al.} 1999), 
methods of characterizing
the orbits can be extremely useful, especially when an attempt is made to understand what the 
``typical'' behavior is of a very large collection of orbits in a particular physical system. 
These methods, some of which are discussed below in \S \ref{char},
are most useful when they are quantitative, reliable, and applicable to a wide variety of systems.  
\\
\indent
Here, the analysis will focus on stellar orbits in models of time-invariant galactic potentials. 
With this in mind, a brief review of some basics of stellar dynamics is presented.  Stars in 
galaxies are presumed to form a collisionless
system, that is, the forces felt by any one star are due only to the mean forces produced by
all other stars.  The distribution function, $f$, of stars must therefore satisfy the collisionless
Boltzmann equation \citep{bnt87},
\begin{equation}\label{boltz}
\frac{df}{dt}=\frac{\partial f}{\partial t} + \sum_{i} \frac{\partial f}{\partial w_i} \dot{w_i}=0,
\end{equation}
where the $w_i$ are the phase space coordinates, e.g., ($x,y,z,\dot{x},\dot{y},\dot{z}$).  A 
by-product of this equation is that any time-independent function of phase space coordinates, known 
as an integral of motion, must be a 
valid solution of eq. (\ref{boltz}).  This fact is incorporated into the Jeans Theorem: any 
steady-state solution of eq. (\ref{boltz}) must be a function of integrals of motion and any 
function of integrals of motion must likewise solve eq. (\ref{boltz}) \citep{bnt87}.  A stronger,
qualified statement can be made.  If all orbits in a given potential are regular\footnote{We use
the term ``regular'' to describe orbits that respect a number of isolating integrals of motion
that is greater than or equal to the degrees of freedom of the orbit \citep{bnt87}.  For 
example, an orbit
in a 2D potential must have at least two isolating integrals in order to be considered regular.
On the other hand, we use ``quasi-ergodic'' 
as a blanket term for both stochastic and semistochastic orbits; irregular orbits are fully 
ergodic.  Ergodic and quasi-ergodic orbits respect only one complete 
isolating integral of motion, either the 
specific energy, $\epsilon$, or the Jacobi constant, $\epsilon_J$.}
then any solution to eq. (\ref{boltz}) is a function of three integrals of motion.
This is known as the Strong Jeans Theorem \citep{bnt87}.  Indeed, in potentials that have
analytically prescriptible integrals of motion, such as spherical or axisymmetric systems, a
variety of distribution functions have been investigated (e.g., King 1966; Kalnajs 1976). 
\\
\indent
However, most realistic models of galaxies are not as simple as those discussed
above, and different tactics must be used to investigate the properties of the stellar 
distribution functions that are associated with these more realistic systems.  And while an
in-depth discussion of the creation of such models is beyond the scope of this paper, a short
overview is given here.
N-body simulations (e.g., Miller \& Smith 1979; Hohl \& Zang 1979;
Combes \& Sanders 1981; Miller {\it et al.} 1982; Pfenniger \& Friedli 1991) have been widely 
used to generate stellar distribution functions for steady-state galaxy models.  By design, many 
of these simulations have
begun with axisymmetric distributions of point particles, then the dynamical systems quickly deform
to bar-shaped objects.  Most of these studies have focused on the global evolution of the
bar, but some \citep{mns79,pnf91} have also looked at constituent orbits of the bars and have
found evidence for nonanalytical integrals of motion.  Performing numerical integrations of 
individual orbits in analytical potentials provides further evidence for these nonanalytical
integrals in studies designed to find periodic orbits (e.g., Heisler \textit{et al.} 1982; 
Magnenat 1982; Mulder \& Hooimeyer 1984; Pfenniger 1984; Martinet \& de Zeeuw 1988; Hasan 
\textit{et al.} 1993).  Uncovering periodic orbits has been very important in the context of
studies of 3D galaxy models because \citet{schwarz79,schwarz82} developed procedures for creating 
self-consistent potential-density pairs (and associated steady-state distribution functions) once 
the periodic orbits for a given 3D potential are known.  These techniques have been extended to
model realistic stellar dynamical systems (e.g., Rix {\it et al.} 1997; van der Marel {\it et al.}
1998; Cretton {\it et al.} 1999; Cretton, Rix, \& de Zeeuw 2000).  
\\
\indent
While the importance of regular orbits has been the focus of
many stellar dynamics research projects, the importance and likely existence of quasi-ergodic orbits 
in realistic galaxy potentials also has been discussed \citep{gns81,bin82b,hab97,vnm98}.
While the previously mentioned studies involved analytical potentials, \citet{us00} have also 
found that quasi-ergodic stellar orbits are quite common in a numerically-created, gaseous bar. 
Quasi-ergodic orbits also appear in N-body simulations of galactic bars (see, for example,
Sparke \& Sellwood 1987). 
A specific example of a situation in which quasi-ergodic orbits play an important role is in
galaxy models with massive central objects (e.g., Udry \& Pfenniger 1988, Valluri \&
Merritt 1998).  In a detailed analysis of the influence that massive central objects have on 
stellar orbits, \citet{gnb85} suggest that sufficiently massive central objects
will scatter into quasi-ergodic orbits trajectories that initially: 1) are regular; 2) pass 
near the central object; and 3) support nonaxisymmetric 
distributions, such as bars.  With the current interest in
understanding active galactic nuclei as well as the recent discovery of a link between the
masses of central objects and central stellar velocity dispersions \citep{fnm00,geb00a}, more 
accurate modeling of the distribution functions of such systems (along the lines of van der Marel
{\it et al.} 1998) could provide some insight into the observational evidence.
\\
\indent
In an effort to better characterize orbits (both regular and quasi-ergodic) in astrophysically
interesting potentials, the relative utility of the correlation integral method has been examined.  
As pointed out by \citet{gnp83} and \citet{cns84}, the correlation integral method is a 
flexible and accurate characterization technique
that has been widely utilized by physicists but has been virtually ignored by the 
stellar dynamical community.  This technique distinguishes orbits based on the number of isolating 
integrals of motion that are respected by any
given orbit.  This makes it a useful, quantitative tool for examining stellar distribution
functions that would arise in potentials such as
the ones mentioned above.  Regular and nonregular orbits can be differentiated.  Within
the regular class, periodic and quasi-periodic orbits can also be distinguished from one
another.  Additionally,
this technique can distinguish between 3D orbits that respect two isolating integrals and 
those that respect only one.  This ability to distinguish in a clear and quantitative fashion
between various types of orbits makes the correlation
integral method particularly well suited to involvement in studies of 3D potentials where 
quasi-ergodic orbits are likely to play an important role.  The primary objective of this paper
is to demonstrate the quantitative reliability and broad utility of the correlation integral method
by applying it as a characterization tool to orbits in a wide variety of potentials -- most of 
which are familiar to the stellar dynamics community -- and by comparing and contrasting it to
other techniques that have been broadly used to characterize orbits.  This will place the community 
in a position to effectively utilize this tool in a wide variety of problems.
\\
\indent
The remainder of this paper is divided as follows.  The different models (one map and several 
potentials) are introduced individually in \S 2.  Section 3 contains summaries of the numerical
integration technique and various orbital characterization methods that are utilized in this study.  
Tests of and results acquired with the correlation integral method are presented 
in \S 4.  Also discussed in this section are the results from a previously unexamined 3D
potential.  Finally, \S 5 presents a summary of the findings.
\\
\section{Models}\label{models}
\indent 
In order to illustrate the basic utility of the correlation integral method, as well as
the accuracy of this particular implementation of the method, several systems that previously 
have been well studied and characterized using other techniques are presented.
\\
\subsection{H\'{e}non Map}\label{henmap}
\indent
The simplest system to attack using the correlation integral method is one that can be contained 
in a 2D phase space.  The H\'{e}non map is one such system that exhibits nontrivial behavior.
It is prescribed by the following set of iterative equations \citep{hen76}:
\begin{equation}\label{henx}
x_{i+1}=y_i + 1 -a x_i^2,
\end{equation}
and
\begin{equation}\label{heny}
y_{i+1}=bx_i,
\end{equation}
where $a$ and $b$ are constants.  Rather than solving a pair of
coupled differential equations for position and velocity, phase space is occupied by
choosing initial $x$ and $y$ values, then iterating back and forth between these two
algebraic expressions.  When $a=1.4$ and $b=0.3$, the H\'{e}non map has a strange 
attractor in phase space, that is, the dimensionality of occupied phase space is noninteger 
(fractal).  
\\
\subsection{Richstone Potential}\label{richpot}
\indent
Following the lead of \citet{cns84}, orbits in the scale-free, logarithmic 
potential referred to here as the Richstone potential \citep{rich80,rich82} are also studied.
(A triaxial version of this potential has been studied in Binney 1981 and is known as 
Binney's potential.)
One nice feature of this potential is that it can be used to study either 2D or
3D orbits (4D and 6D phase spaces, respectively).  To investigate fully 3D orbits, 
the following form of the potential is utilized,
\begin{equation}\label{3drich}
\Phi_{\rm R}(x,y,z)=\frac{v_0^2}{2} \ln\left(x^2 + y^2 + \frac{z^2}{q^2} + R_c^2 \right),
\end{equation}
where $v_0$ is the constant circular speed for the potential, $q$ is a measure of the 
flattening of the potential, and $R_c$ is a core radius.  There are three possible 
potential shapes:
\begin{itemize}
\item $q>1$ -- the potential is prolate spheroidal;
\item $q=1$ -- the potential is spherical;
\item $q<1$ -- the potential is oblate spheroidal.
\end{itemize}
As in \citet{rich82}, the parameters are set as $q=0.75$ and $R_c=0.1$.
\\
\indent
In order to examine 2D orbital motion in the Richstone potential, thereby reducing the
analysis from a 6D to a 4D phase space, first transform eq.(\ref{3drich})
to cylindrical coordinates
$(R,\phi,z)$.  It is clear that $\Phi_{\rm R}$ is axisymmetric, so
the Lagrangian is cyclic in $\phi$, which means that the $z$-component of angular 
momentum is constant.  With this restriction, orbits in the 2D Richstone potential 
move under the influence of the following effective potential,
\begin{equation}\label{2drich}
\Phi_{\rm R,eff}(R,z)= \frac{v_0^2}{2} \ln\left(R^2 + \frac{z^2}{q^2} + R_c^2 \right) + 
\frac{L_z^2}{2R^2}. 
\end{equation}
It is this 2D potential that \citet{cns84}
used to validate their implementation of the correlation integral method. 
\\
\subsection{H\'{e}non-Heiles Potential}\label{henheil}
\indent
The H\'{e}non-Heiles potential \citep{hnh64} is a well-known potential that supports both
regular and quasi-ergodic orbits.  This potential has the form,
\begin{equation}\label{henonpot}
\Phi_{\rm HH}(R,z)= \frac{1}{2}(R^2+z^2+2R^2z-\frac{2}{3}z^3).
\end{equation} 
For energies $\epsilon \lesssim 0.01$, 
almost all orbits in this potential are regular.  As the energy is increased, more and 
more of phase space is occupied by quasi-ergodic orbits.  At an energy $\epsilon = 1/6$, 
the situation is reversed in that most orbits are quasi-ergodic.
\\
\subsection{St\"{a}ckel Potential}\label{stack}
\indent
Since St\"{a}ckel potentials are separable, all orbits in such potentials are regular.
This feature makes orbits in St\"{a}ckel potentials ideal subjects for testing characterization
techniques.  The St\"{a}ckel potential that is used in this study has the form \citep{dez85},
\begin{equation}\label{stackpot}
\Phi_{\rm S}=\frac{-v_0^2}{1+(\frac{x}{a})^2 + (\frac{y}{b})^2 + (\frac{z}{c})^2},
\end{equation}
where $v_0=1.0$, $a=1.0$, $b=0.8$, and $c=0.6$.  This is a triaxial potential where the minor
axis lies along the $z$-axis, the intermediate axis lies along the $y$-axis, and the major
axis lies along the $x$-axis.
\\
\subsection{Cazes Bar Potential}\label{cbarpot} 
\indent
In an effort to investigate the fission hypothesis for binary star formation, \citet{caz99}
has produced two models of rapidly rotating, steady-state, triaxial, gaseous density
distributions using a numerical hydrodynamics code.  These models have compressible
($n=3/2$) polytropic equations of state and nontrivial internal flows, including two 
relatively weak standing shocks.  A detailed description of these models appears in 
\citet{cnt00}.  The hydrodynamics simulation that created these models was performed with
dimensionless units.  This allows the results to be scaled to a variety of
systems simply by choosing an appropriate mass and length scale \citep{wnt87}.  For example, 
choosing a
mass of $1.0\, M_{\odot}$ and a length scale $R_{\rm eq} \approx 4.57 \, {\rm AU}$ describes a 
protostellar object of density $\rho \approx 10^{-9}\, {\rm g\, cm^{-3}}$ \citep{caz99}.
Alternatively, for a mass of $10^{10}\, M_{\odot}$ and length scale $R_{\rm eq} \approx
2\, {\rm kpc}$, the now galactic-sized object has a density $\rho \approx 10^{-23}\, 
{\rm g\, cm^{-3}}$.  With this galactic scaling in mind, the \citet{cnt00} ``Model
B'' is adopted as an example of a nontrivial, rotating, bar-like structure with a 
self-consistent and realistic potential-density pair.  Hereafter, this model will be referred
to as the ``Cazes bar.''
\\
\indent
Since the Cazes bar
was created numerically, values of various quantities such as mass density and the gravitational
potential are specified at discrete locations on a computational
grid; specifically, \citet{cnt00} used a $128 \times 64 \times 256$ cylindrical $(R,z,\phi)$ 
grid.  For reasons that will be made clear in a later section, it is difficult to accurately
evaluate Lyapunov exponents and the correlation integral on such a coarse, discrete grid.
In analyzing orbits in the Cazes bar potential, an analytical potential 
constructed along the lines discussed in \citet{us00} to closely resemble the numerical Cazes bar 
potential will be adopted instead.  Specifically, the following analytical potential will be 
used:
\begin{eqnarray}\label{acazes}
\Phi_{\rm aCB}(x,y,z) & = &  N\bigg\{ 1- \left( 1 + 
\left( \frac{x}{R_{\rm L2}} \right) ^{\alpha} \right. \nonumber \\ 
 & + &  \left. \left( \frac{y}{qR_{\rm L2}} \right) ^2  + 
\left( \frac{z}{q_z z_{\rm lim}} \right) ^{\gamma}\right) 
^{-\beta} \bigg\} \nonumber \\
  & - & \frac{1}{2}\Omega^2(x^2 + y^2) + \Phi_{\rm min},
\end{eqnarray}
where $N$ is a normalization factor; $q$ determines the strength of the bar-like
distortion in the equatorial plane; $q_z$ determines the strength of the bar distortion 
in the $x-z$
plane; and $\alpha$, $\beta$, and $\gamma$ are exponents whose values are to be 
determined.  $R_{\rm L2}$ is the distance along the major axis from the center of the bar 
to its ${\rm L2}$ 
Lagrange point, and $z_{\rm lim}$ is a vertical scale height.  Only values
of $q$, $q_z$, $z_{\rm lim}$, and $R_{\rm L2}$ for which the $x$-axis coincides with the 
major axis of the bar will be considered; the $y$-axis is then the intermediate axis; and 
the $z$-axis is along the 
bar's minor axis.  The angular velocity of the bar and the value of the potential at 
the center are taken from the numerical Cazes bar, that is $\Omega=0.5218$ and 
$\Phi_{\rm min}=-1.018$, respectively.  When $z$ is set to zero, this potential reduces
exactly to the 2D analytical potential that was studied in \citet{us00}.
\\
\indent
In an effort to illustrate how well the analytical approximation to the Cazes bar potential
$\Phi_{\rm aCB}$ matches the potential that was derived numerically in the \citet{cnt00}
hydrodynamical simulation, Fig. \ref{acbvsncb} presents equipotential contours 
from both the numerical Cazes bar and the analytical Cazes bar.
Figure \ref{acbvsncb}a shows a slice of the numerical Cazes bar along the positive half 
of the major axis in the meridional plane.  The corresponding
slice from the analytical Cazes bar is shown in Fig. \ref{acbvsncb}b.  The parameters for 
the analytical potential pictured here are: $N=0.7$, $q=0.8$, $q_z=1.5$, $R_{\rm L2}=1.36$, 
$z_{\rm lim}=0.65$, $\alpha=4$, $\beta=4$,
and $\gamma=1.7$.
Equipotential contours in the $y-z$ plane of the numerical Cazes bar
are illustrated in Fig. \ref{acbvsncb}c.  Figure \ref{acbvsncb}d displays the corresponding 
analytical Cazes bar
plot.  Figures \ref{acbvsncb}e and \ref{acbvsncb}f display equipotential contours in the 
equatorial planes
of the numerical Cazes bar and analytical Cazes bar, respectively.  The properties of this 
equatorial-plane, 2D potential
were investigated thoroughly in \citet{us00}; it will be used below as one of the test potentials.  
As a check
on the applicability of the analytical Cazes bar as a substitute for the numerical Cazes bar,
several integrations with identical initial conditions in both potentials have been performed.  
In every case, the resulting orbital projections had similar (although not exact) morphologies.
\\
\section{Integration and Characterization Tools}\label{char}
\subsection{Numerical Orbit Integration}\label{verlet}
\indent
Over the past few decades, numerical integrations of the equations of motion,
have become the standard way to investigate stellar orbits in galaxies.  In general, the
equations of motion have the following vector form,
\begin{equation}\label{eqmot}
\ddot{\vec{x}}=-\vec{\nabla}\Phi - [\vec{\Omega} \times (\vec{\Omega}\times \vec{x})] 
- 2\vec{\Omega} \times \dot{\vec{x}},
\end{equation} 
where $\vec{\Omega}$ is the angular velocity of a rotating frame of reference, $\vec{x}$ is
the position vector in that frame, and $\Phi$ is the gravitational potential of the system.
When a rotating potential is investigated in this paper, the angular velocity vector always
points in the $z$-direction.
\\
\indent
Orbits are calculated with a Verlet integration scheme \citep{ver67}.  This is straightforward
to implement for non-rotating potentials, such as the 2D and 3D Richstone potentials.  
When a rotating
potential is examined, the Verlet scheme is modified so that two Verlet steps are 
performed per fixed timestep.  This is done because Coriolis terms must be 
included in the accelerations.  In order to provide optimum values of velocities for evaluation
of the 
Coriolis terms, a first Verlet step is used to obtain a first estimate of the velocities, but
particle positions and velocities are not permanently 
updated at this step.  Then, for the same timestep, the second Verlet step 
returns and updates a more accurate subsequent position and velocity.  This procedure provides 
an adequate level of energy conservation ($\Delta \epsilon/ \epsilon < 10^{-6}$ over $10^9$ 
timesteps) for the purposes of this investigation.  The need for this level of energy 
conservation is the
reason that an analytical formula approximating the Cazes bar potential has been adopted 
in preference to the original, numerically
prescribed $\Phi_{\rm CB}$.  Using an interpolation scheme with the
numerical Cazes bar, the energy error (with the same timestep) is $\Delta \epsilon/ \epsilon 
\approx 10^{-3}$.  While this is adequate for a determination of the overall shapes of 
various orbits, the correlation 
integral method requires that the orbit integration scheme provide much better energy 
conservation.  In principle, the numerical potential could be utilized if a smaller
integration timestep were used.  In practice, though, this made total integration times too 
long to be practical.  This retreat to an analytical form of the Cazes bar potential should not 
be taken as a withdrawal from the assertion that the correlation integral
method is applicable to numerically derived potentials.  On a faster machine (or with an
improved interpolation scheme) the computational costs would not have posed as great a problem
and a numerical potential could have been used.  
\\
\indent  
Initial conditions for the orbits presented in this paper have not necessarily been chosen 
to represent a complete sample of phase space.  Instead, the orbits discussed here illustrate
different (but not all) families that exist in the various potentials.  For example, in the
H\'{e}non-Heiles potential, two regular and two quasi-ergodic orbits are followed to illustrate
their respective differences.  Orbits in the 3D analytical Cazes bar potential are populated
under the Restriction Hypothesis of \citet{us00}.  The idea behind the Restriction Hypothesis is 
that if stars form from gas in a system, the initial velocities of the stars will be determined
by the gas flow at the points of formation.  In practice, the Restriction Hypothesis is imposed 
by choosing initial positions that coincide 
with cylindrical grid points where the gas velocities $(v_{\rm x,gas},v_{\rm y,gas},
v_{\rm z,gas})$ from the numerical Cazes bar are known and most definitely does not
represent a complete sampling of phase space.
\\
\subsection{Orbit Morphology}\label{morph}
\indent
Perhaps the simplest way to categorize any orbit is according to its shape in
configuration $(x,y,z)$ space.  Examples in 2D potentials are
the ``banana'' orbits discussed in \citet{bin82a}, the $4/1$ orbits described in 
\citet{con88}, and the ``bow tie'' orbits presented in \citet{us00}.
For 3D orbits, projections of an orbit onto the three principal planes of the chosen
configuration space coordinate system
can illustrate the orbital morphology and therefore also can be useful when 
categorizing orbits.  For example, if the $x-y$ plane projection of an orbit appears
to have definite circulation about the center (i.e., the $z$-component of the orbital
angular momentum, $L_z$, is never zero), and the
$y-z$ and $x-z$ projections appear as rectangular areas, the orbit is called a $z$-axis
tube orbit \citep{bnt87}.  The benefit of characterizing orbits in this way is that such a 
descriptive name confers a basic image of the orbit shape that can be recognized in other
contexts.  The drawbacks to this method
are that it is descriptive rather than quantitative; complex orbits usually cannot be 
simply defined in words; and sometimes the adopted terms
are not universally accepted (e.g., Richstone 1982).  Another possible problem that
arises when dealing with 3D orbits is that, while one planar projection may be simply described, 
the other two complementary projections may be too complex for words (see, for example, Fig. 9 
in Pfenniger 1984).  
Although orbital morphology will not be used as a primary characterization tool in this 
paper, orbit projections will be utilized as visual aids when discussing most orbits.
\\
\subsection{Spectral Dynamics}\label{spectral}
\indent
Another way of classifying 2D and 3D orbits is the spectral dynamics method first
introduced to the astrophysical community by \citet{bns82}.  Briefly, the spectral dynamics 
characterization method
is based on the fact that regular orbits in near-integrable potentials exist in regions of 
phase space with toroidal topology.  
As noted earlier in footnote 1, here the term ``regular'' means that the number of isolating 
integrals respected by the orbit is greater than or equal to the dimensionality of the orbital 
configuration space.
These phase space tori can be described in terms of actions and action angles, so the Fourier 
spectra of the $(x(t),y(t),z(t))$ components of a regular orbit must consist of discrete lines.
The frequencies of these lines are linear combinations of the fundamental frequencies
associated with the action angles.  The exact linear combinations of frequencies contain
information about the resonances of the orbit.  Since ergodic and quasi-ergodic orbits do not 
lie on tori in phase space, their spectra will consist of a forest of lines at all 
frequencies.  For more complete discussions of spectral dynamics, the reader is directed to 
\citet{bns82}, \citet{lask93}, \citet{cna98}, and \citet{copet00}.  
\\
\indent
Despite the demonstrated utility of the spectral dynamics method in a variety of contexts, 
it has not been incorporated into the present study for two reasons.
Primarily, the spectral dynamics method does not provide a quantitative measure of 
quasi-ergodic orbits whereas, as
mentioned in the introduction, realistic galactic potentials probably support quasi-ergodic 
orbits.  Secondly, the spectral dynamics method is strictly applicable only to
``near-integrable'' systems.  This means that the Hamiltonian for a given system must be 
expressible in terms of an integrable part plus some small perturbation.  Numerically 
determined potentials, or their analytical approximations, generally cannot
be expected to fall into this category.  
\\
\subsection{Lyapunov Exponents}\label{lyapintro}
\indent
A more quantitative description of an orbit can be made using Lyapunov exponents.  For a
given phase space trajectory, $\vec{\xi}(t)=(x(t),y(t),z(t),p_x(t),p_y(t),p_z(t))$, the
Lyapunov exponents measure how a nearby trajectory, $\vec{\xi}(t) + \vec{\Delta \xi}(t)$, 
where $\vec{\Delta \xi}$ is initially small, diverges from the given phase space orbit.  
A number of previous studies of orbits in an astrophysical context have utilized Lyapunov
exponents to characterize the orbits.  A
few examples that contain excellent discussions of the technique are \citet{mnv96},
\citet{hab97}, and especially \citet{unp88}.  Briefly, there is one Lyapunov exponent, $k_i$,
for each dimension $i$ of phase space.  In conservative (Hamiltonian) systems (such as 
all the cases discussed in this paper),
these exponents are not independent.  In fact, $\sum_i k_i =0$.  Also, the restriction 
that phase space volumes must remain constant (Liouville's theorem) connects conjugate pairs
of exponents.  In addition, the exponent that corresponds to motion along the direction
of motion is zero.  For example, in a 4D phase space two exponents are equal to zero
(one is along the direction of motion and the other is in the conjugate direction); the
other two exponents are equal in magnitude but have opposite signs \citep{lnl83}.  In
particular, what is measured and referred to as the Lyapunov exponent in this paper
is the value of the largest positive exponent.  The attractive attributes
of this method are its quantitative nature and its applicability to orbits in any
potential.  The main drawback to using this method as a solo
characterization tool is that no distinction can be made between closed and unclosed regular
orbits.  That is, all regular orbits display insensitivity to small changes in initial 
conditions and therefore show similar behaviors in their Lyapunov exponents.
\\
\indent
The technique for measuring the largest Lyapunov exponent follows from the prescription 
given in \citet{ben76}.  \citet{lnl83} give
a good introduction to the method, starting from the definition of the Lyapunov exponent,
which is summarized here.
The assumption is made that two nearby trajectories diverge exponentially with time, i.e.,
\begin{equation}
d(t)=d(0) e^{kt},
\end{equation}
where $d$ denotes a phase space distance.  The Lyapunov exponent can then be defined to be,
\begin{equation}\label{lyapk}
k \equiv \lim_{t \to \infty, d(0) \to 0} \left(\frac{1}{t}\right) \ln \left[\frac{d(t)}{d(0)}
\right].
\end{equation}
If a chosen orbit is quasi-ergodic, so that two nearby trajectories diverge as $e^{\sigma t}$, 
then $k=\sigma =$ constant.
If, instead, the chosen orbit is regular and nearby trajectories diverge only as a power law 
in time, $d(t) \sim d(0) t^{\alpha}$, then 
\begin{equation}
 k=\frac{\alpha}{t} \ln t.  
\end{equation}
In this case, a plot of $\ln k$ vs. $\ln t$ should have a slope $\approx -1$ for large 
values of $t$, independent of the precise value of the exponent $\alpha$.
\\
\indent
The definition of $k$ as given in eq.(\ref{lyapk}) is computationally unsatisfactory.  
Exponential growth can quickly lead to numbers that a computer cannot represent.  \citet{ben76} 
suggest, instead, that an orbit be broken into $n$ finite time lengths, $\tau$ (see 
their Fig. 1 or Fig. 5.6 in Lichtenberg \& Lieberman 1983 for a pictoral representation of 
this idea).  Then, every $\tau$
time units, the distance between neighboring trajectories should be re-normalized to the distance 
between the two at the beginning of the orbit.  With this technique, the Lyapunov exponent
is defined to be \citep{ben76},
\begin{equation} 
k_n=\frac{1}{n\tau} \sum_{i=1}^{n} \ln \frac{d(i\tau)}{d(0)},
\end{equation}
where, in the limit of $n \to \infty$, $k_n \to k$.
\\
\indent
The task that remains is to calculate whether or not nearby trajectories exponentially diverge.
The vector of interest here is $\vec{\Delta \xi}$, that is, the vector difference between the orbit 
that is integrated and the nearby trajectory.  Following \citet{lnl83}, define 
$\vec{w} \equiv \vec{\Delta \xi}$.  From a linear stability analysis (as in Binney \& Tremaine
1987, \S 3.5.3), then, 
\begin{equation}\label{weqns}
\frac{d\vec{w}}{dt}= \mathcal{M} \cdot \vec{w},
\end{equation}
and $\mathcal{M}$ is a tensor that has components defined by,
\begin{equation}
\mathcal{M} \equiv \frac{\partial \vec{F}}{\partial \vec{\xi}},
\end{equation}
where 
\begin{math}
\vec{F}=\left( -\frac{\partial H}{\partial \vec{x}}, \frac{\partial H}{\partial \vec{p}}
\right),
\end{math}
$H$ is the system's Hamiltonian, $\vec{x}$ is the generalized coordinate vector, and 
$\vec{p}$ is the conjugate momentum vector.  In a system with $N$ degrees of freedom, there
are now $3N$ equations that must be solved -- $N$ for the particle trajectory and $2N$ for the
phase space difference vector.
\\
\indent
The $2N$ differential equations given by eq.(\ref{weqns}) are translated into finite-difference
equations that are solved alongside the equations of motion.  
For orbits that will be 
analyzed in this paper, $k_n$ is determined for $n\tau =10^2$, $5\times 10^2$, $10^3$, 
$5\times 10^3$, and $10^4$.  In addition, the value of $\tau$ has been chosen to be the
same as $\Delta t$, so the difference vector is renormalized every timestep.  The 
presentation of Lyapunov exponents is in the form of plots showing $\ln k_n$ versus 
$\ln n\tau$.  This form has been chosen for these plots because regular orbits (those 
insensitive to small 
changes in initial conditions) have slopes $\approx -1$, while quasi-ergodic orbits have 
slopes $\approx 0$ because they are sensitive to initial conditions.
\\
\subsection{Correlation Integrals}\label{introcorr}
\indent
The main orbital characterization method that will be utilized in this paper is the 
correlation integral method.  As described by \citet{gnp83} and \citet{cns84},
the correlation integral provides a measurement of phase space dimensionality.
The basic idea underlying this method is that for a given orbit, the correlation 
integral is calculated for all phase space.  Then, since a plot of 
the correlation integral versus phase space distance behaves like a power law
for small phase space distances \citep{gnp83}, the exponent of the power law provides a 
measure of the dimensionality
of the phase space that is occupied by the orbit.  Once this dimensionality is known, one can
also readily deduce the number of isolating integrals that are respected by the selected orbit.  
Approaching a previously unexamined potential in this way gives, for example, basic information
about whether or not most orbits are regular.  This characterization then
may or may not lead one to perform a more complex analysis, such as spectral dynamics.  
\\
\indent
Operationally, the steps in determining the dimensionality of a phase space orbit are:
\begin{enumerate}
\item Integrate an orbit for a sufficient number of timesteps (``sufficient'' will be 
      clarified below).
\item Choose a set of sampling points from the orbit.
\item Calculate the correlation integral, $C(r)$, as defined by eq. (\ref{corrint}).
\item Measure a slope from a plot of $\ln C(r)$ vs. $\ln r$.
\end{enumerate}
\indent
Because topics 1 and 2 are somewhat connected, they are discussed together.  From 
various trials, investigating phase spaces of different sizes, it has been determined that 
approximately 50 orbital periods must be completed in order for an accurate correlation
integral to be found.  This means that more timesteps are necessary for each 3D orbit than
for each 2D orbit.  With the timestep that has been used throughout this investigation, the 
ratio between the number of required timesteps is $\approx 100$.
(It is likely that these numbers would change if a different integration scheme was used.) 
The sampling points
are chosen following the suggestion in \citet{cns84}; specifically, random points are selected 
from a
subset of the orbital phase space points.  The subset consists of $10^5$ points taken at 
equal timestep
intervals.  Normally, $10\%$ of these points are randomly chosen to be sampling points.  
While this
number of sampling points is usually adequate for the purposes of this study, 
it generally is best to use as many sampling points as is computationally practical. 
\\
\indent 
The correlation integral is determined numerically using the following formula 
\citep{gnp83}:
\begin{equation}\label{corrint}
C(r)= \lim_{N \to \infty}\frac{1}{N^2}
\sum^N_{i=1} \sum^N_{j=1,\neq i} \Theta(r - |\vec{\xi_i} - \vec{\xi_j}|),
\end{equation}
where $\Theta$ is the Heaviside step function, the $\vec{\xi_i}$ are phase space position 
vectors, $r$ is a phase space distance, and
$N$ is the number of sampling points from the orbit.  It is now clear why it is 
advisable to use
as large a value of $N$ as practical, as the correlation integral is defined more accurately 
when $N$ is larger.  A naive evaluation of eq.(\ref{corrint}) is very slow for large $N$, even 
if one takes advantage of the symmetry of the Heaviside function.  Dividing phase space into
``bins'' can significantly speed up the evaluation of the double summation.
\\
\indent
The final step in determining the dimensionality of a phase space orbit is measuring a 
slope from the plot of $\ln C(r)$ vs. $\ln r$.  A log-log plot is used because of the 
behavior of the
correlation integral.  \citet{gnp83} state that, for $r \ll 1$, $C(r) \propto r^{\nu}$,
where $\nu$ is the dimensionality of the phase space orbit.  So, when $\ln C(r)$ 
is plotted against $\ln r$, the slope of any
linear section can be interpreted as the dimensionality $\nu$ of the phase space orbit within 
that range of 
$r$.  In order to ascertain the reliability of the values of $\nu$ that have been measured for
individual orbits, a minimized $\chi^2$ linear fit to the $\ln C(r)- \ln(r)$ 
data for five independent sets of sampling points has also been calculated.  Also, at least one 
order of magnitude in $r$ 
must be covered by the linear section to be considered.  It is the average slopes ($D=\langle
\nu \rangle$) and 
standard deviations ($\sigma$) of these five linear fits that are reported in the legends of
figures such as Fig. \ref{2dcazes1}b.  
\\
\section{Results}\label{result}
\indent
The individual phase space orbits that are discussed in this section are listed in Table 1.
As indicated in the first column of Table 1, the orbits
are identified by the potential (or mapping) in which they exist.  Listed in column 2
are the corresponding figure numbers that contain visual summaries of the orbital analysis.  The 
initial conditions (positions, velocities, and energies) for each of the orbits are listed in the next 
seven columns.  The last two columns of Table 1 are provided as a quick reference to the Lyapunov 
exponent and correlation integral characterizations that have been determined for each orbit.  If the 
next to last column (labeled Lyapunov) contains 
an `I' (denoting an orbit insensitive to small changes in initial conditions), that orbit is 
regular;  an `S' (for sensitive) denotes an orbit that is not regular.  The last column holds 
the measured dimensionality $D$ of each phase space orbit.  
\\
\indent
Before undertaking a discussion of the characterization of these selected orbits, let us examine
the expectations for the various models.  The 2D phase space of the H\'{e}non map is not derived 
from a Hamiltonian system, so there are no integrals of motion for this system {\it per se}.  
However, each of the 4D phase spaces (2D Richstone, H\'{e}non-Heiles, and 2D Cazes bar) is 
a Hamiltonian system.  In these cases, the potentials are time-invariant
so the energy (or $\epsilon_J$) must be an integral, and we should find 
that $D \le 3$ for all orbits.  Fully ergodic orbits will have $D = 3$, while quasi-ergodic orbits
will have $2 < D < 3$.  Quasi-ergodic orbits respect only one full integral of motion, but they also
have some (unknown) restriction in addition to that of their ergodic cousins.  Both ergodic and 
quasi-ergodic orbits are sensitive to initial conditions, so the Lyapunov exponent
should remain nearly constant with time, as discussed in \S \ref{lyapintro}.  
On the other hand, a regular orbit must have a total 
number of isolating integrals that is equal to or greater than the number of spatial degrees of 
freedom (in this case $\ge 2$; specific energy plus 1 or 2 unknown integrals).  
So, the allowed phase space for each regular (but not periodic) orbit should be 2D (4 phase 
space dimensions 
minus the 2 isolating integral dimensions).  The behavior of the Lyapunov exponent should
be that of a regular orbit, that is, the slope of $\ln k_n$ vs. $\ln n\tau$ should be
$\approx -1$.
\\
\indent
Orbits in 3D Hamiltonian systems occupy 6D phase spaces.  However, since the potentials are
still time-invariant, the measured dimensionality 
of the phase space orbit should be, at most, five.  Regular orbits in 3D potentials must
respect at least three integrals of motion, $D \le 3$, and as before, the Lyapunov exponent
should decrease with time for a regular orbit.  Returning to the nonregular orbits, fully
ergodic orbits should display a dimensionality $D = 5$, while quasi-ergodic orbits should have 
a noninteger value $4 < D < 5$.  Orbits with $3 < D \le 4$ are neither regular nor quasi-ergodic.  
They respect (at least) two full integrals
of motion and will be referred to as semiregular orbits in this paper.  All nonregular orbits
have Lyapunov exponent plots that have $\ln k_n$ vs. $\ln n\tau$ slopes that are $\approx 0$.
\\
\subsection{2D Phase Space}\label{2dps}
\indent
As mentioned earlier, the H\'{e}non map phase space is fractal (see \S\ref{henmap}).  The 
accepted value for the dimensionality of this phase space orbit is $D=1.25 \pm 0.02$ 
\citep{gnp83}.  
Figure \ref{henplot}a shows the $(x,y)$ phase space structure of the H\'{e}non map; Fig. 
\ref{henplot}b is a magnified view illustrating the fractal nature of this phase space orbit.
Figure \ref{henplot}c 
shows the $\ln C(r)$ vs. $\ln r$ plot (hereafter, the $C-r$ plot) that was obtained in this study 
for the H\'{e}non map using 
the correlation integral method.  The result is $D=1.21 \pm 0.01$.
This is also the value that \citet{gnp83} obtained using 
the correlation integral method.  The difference between this result and the accepted value 
is discussed in \citet{gnp83b}.  They claim that
there is a systematic error in the determination of the H\'{e}non map correlation integral.
Interestingly, this error does not occur in correlation integral determinations for other maps
such as the Kaplan-Yorke map, the logistic equation, or the Lorenz equation \citep{gnp83b}.  They
conclude that the error in the H\'{e}non map result arises from the sample phase space points
not uniformly covering the phase space orbit.
However, their suggested remedy does not work well when phase space is populated by numerically
integrated orbits.  In the following sections it will be shown that this slight disagreement does
not diminish the quantitative usefulness of the correlation integral method as a tool for
characterizing orbits.
\\
\subsection{4D Phase Space}\label{4dps}
\indent
For orbits in 2D potentials, the presentation of results will adhere to the following form:
In each figure, the frame labeled (a) contains the orbital trajectory; frame (b) displays 
the $C-r$ plot 
obtained using the correlation integral method; frame (c) shows the
Lyapunov exponent plot as $\ln k_n$ vs. $\ln n\tau$; and frame (d) contains either the
($x-p_x$) or ($R-p_R$)
surface of section diagram for the orbit.  The legends of frames labeled (b) also will
provide quantitative information derived from the $C-r$ plot as discussed in \S \ref{introcorr}, 
namely the slope $D$ and 
standard deviation $\sigma$.  In most Lyapunov exponent plots, a dot-dashed line with a slope
of -1 also is included because, as discussed in \S \ref{lyapintro}, the behavior of $k_n$ should
approach this slope as $n\tau \to \infty$ if an orbit is regular.
\\
\indent
The results for four orbits in the 2D Richstone potential are shown in Figs. \ref{2drich1}
through \ref{2drich4}.  
In every case, all three techniques for characterizing the orbits
indicate that the orbits are regular: the surface of section plots are composed of invariant
curves; the Lyapunov exponent drops with time with a slope of minus one; and from the
correlation integral, the measured dimensionality has a value $\le 2$.
Note that the Lyapunov exponent plots in Figs. \ref{2drich1}c and \ref{2drich4}c are
basically identical and therefore make no distinction between the closed and unclosed regular 
orbits.
However, the difference is clearly illustrated by the differing slopes of the $C-r$ plots in
Figs. \ref{2drich1}b and \ref{2drich4}b: the periodic orbit displays $D=1$, instead of $D=2$. 
\\
\indent 
At this point, it is worth noting some general features of $C-r$ plots
derived from orbits (of any dimensionality) as opposed to mappings.  As the value of $r$ 
increases, the plotted points will often deviate from the line of slope $D$.
This is because a clean linear relation is expected only for $r \ll 1$.
Another way to think about this is that, for a sufficiently large value of phase space 
distance $r_0$, the entire phase space orbit will be 
enclosed.  Then, for $r \ge r_0$, $C(r)=1.0$, so the $C-r$ plot must tend toward a slope of zero
at large enough values of $r$.  
Deviations from a linear slope of $D$ may also occur at the smallest values of $r$, but for a 
different reason.  
Since only a finite number of sampling points is used, there is necessarily a
lower limit to the smallest distance that can be measured between any two points.  Basically, 
this is a small-number statistics
problem.  As a larger number of sampling points is used, the linear fit generally becomes 
tighter and extends to smaller values of $r$.
\\
\indent
Four orbits with $\epsilon = 1/8$ in the H\'{e}non-Heiles potential are displayed and
analyzed in Figs. \ref{henh1} through \ref{henh4}.  
Unlike the 2D Richstone potential, the 
H\'{e}non-Heiles potential supports some quasi-ergodic orbits at this energy.  All three
methods of characterization indicate that
the orbits shown in Figs. \ref{henh1} and \ref{henh2} are regular, while the
ones shown in Figs. \ref{henh3} and \ref{henh4} are quasi-ergodic.  Focusing on the two
quasi-ergodic orbits, notice that, (1) the Lyapunov exponent is approximately constant,
signaling an exponential departure of two initially neighboring trajectories; (2) the
surface of section is no longer composed of a smooth invariant curve; and (3) the $C-r$
plot identifies a dimension greater than two.  We know that the orbits 
shown in Figs.
\ref{henh3} and \ref{henh4} are quasi-ergodic rather than fully ergodic because of the
measured dimensionality.  This is a distinction that can be made clearly from the $C-r$
plot, but it is not possible from a measurement of the Lyapunov exponent alone.
The deviations from $D$ at small values of $r$ in Figs.
\ref{henh3}b and \ref{henh4}b are simply artifacts of the limited number of orbital timesteps.  
This is similar to the small number statistics problem that was discussed above.  
When the number of timesteps is increased, these deviations disappear and the line with slope 
$D$ extends to smaller $r$ values.
\\
\indent
Four different orbits that are supported by the 2D analytical Cazes bar potential are 
presented here in Figs. \ref{2dcazes1} through \ref{2dcazes4}.  
Once again, all three methods
of characterizing these 2D orbits agree: the orbits shown in Figs. \ref{2dcazes1}, \ref{2dcazes2},
and \ref{2dcazes3} are regular, while the one shown in Fig. \ref{2dcazes4} is quasi-ergodic.
The orbit shown in Fig. \ref{2dcazes1}a has $D \approx 1$, rather than $D=2$; hence, it is periodic.  
This orbit is the parent of what has been called the `bow tie' family of orbits in \citet{us00}.
The orbit shown in Fig. \ref{2dcazes2}a appears to be trapped near the periodic bow tie orbit.  
The most interesting aspect of this orbit is visible in the $C-r$ plot, Fig. 
\ref{2dcazes2}b.  There are two linear sections in the $C-r$ plot.  The one that exists 
for small $r$
has a slope of $D\approx 2$.  At larger $r$, however, the linear section has a slope of 
$D\approx 1$.  This slope discontinuity persists
even when this orbit is followed through 100 times as many timesteps.  This suggests that 
the discontinuity
and the $r$ location of the discontinuity are physically relevant (as opposed to the cases
in the previous paragraph where differing slopes were numerical artifacts).  We interpret 
this difference using the following analogy presented by \citet{gl87}. 
Imagine viewing a ball of string from very far away.  If asked to describe the dimensionality 
of the ball of string, which appears to be a point, the answer would be zero.  Moving closer, 
the ball can be seen to be extended.  The ball now appears to have dimensionality equal to 2.  
Moving even closer,
the ball now appears to have a finite extent in a direction perpendicular to the dimensions
already present as well, and is thus a 3D object.  However, observing the ball of string
at very close range, the one-dimensional nature of the string becomes apparent.  So, the 
dimensionality of the phase space orbit can change depending on what length scale is 
observed.  Indeed, for small enough values of $r$, all
phase space orbits derived from particle trajectories are really 1D objects.
With this in mind, the segment with slope $D\approx 1$ is interpreted as demonstrating that 
the orbit is not far from being closed.  
The $r$ value at which the change in slope occurs identifies a critical length scale for 
this orbit that cannot be determined via the Lyapunov exponent method or from surfaces of section.  
\\
\indent
The quasi-ergodic orbit shown in Fig. \ref{2dcazes4} has a dimensionality $D\approx 2.54$.
Unlike the discontinuity present in the regular orbit $C-r$ plot (Fig. \ref{2dcazes2}b), 
the apparently linear slope at smaller $r$ values is dependent on the number of timesteps 
taken for the orbit, as with the orbits shown in Figs. \ref{henh3} and \ref{henh4}. 
\\
\subsection{6D Phase Space}\label{6dps}
\indent
As with the 2D orbits previously discussed, the figures containing results for individual 3D 
orbits all have a similar form.  Each figure consists of: a frame labeled (a) illustrating
the $x-y$ projection of the orbit; a frame labeled (b) illustrating the $x-z$ orbital 
projection; a frame labeled (c) illustrating the $y-z$ projection of the orbit; a frame
(d) showing the $C-r$ plot for the orbit; and finally, a frame labeled (e) that displays
the Lyapunov exponent plot derived from the orbit.  
\\
\indent
Figures \ref{3drich1} through \ref{3drich3} display results for three different orbits
in the 3D Richstone potential.  
As expected, these results support the characterization of
these orbits as regular.  In this case, the three integrals of motion are the energy, the
$z$-component of angular momentum (which was explicitly conserved in the 2D case), and the
unknown integral that was also apparent in the 2D case.  These figures demonstrate that
the correlation integral method agrees with other characterization techniques for 3D as
well as 2D orbits.
\\
\indent
While it has been important to use individual orbits to compare characterization methods, it 
is useful to investigate ensembles of
orbits using only correlation integrals.  As stated in \S \ref{stack}, St\"{a}ckel potentials 
provide an excellent test bed for characterization methods.  One hundred orbits of varying energies
(25 orbits each at $E=-0.2$, $-0.4$, $-0.6$, $-0.8$) have been integrated in the potential given
by eq. (\ref{stackpot}).  These orbits have zero initial velocities and initial positions that cover
one octant of the given energy surface.
\\
\indent
The results are shown in Fig. \ref{stackhist}.  
This histogram shows the number of orbits versus
the number of integrals of motion respected by those orbits.  Note that the number of integrals of
motion $I$ is determined by subtracting the $D$ value taken from a $C-r$ plot from the phase space
dimension.  In this case, $I=6-D$.  There is some scatter about the expected value of $I=3$.  This
scatter is due to the fact that only 10,000 sampling points have been used.  For example, the
orbit that here appears to respect only 2.6 integrals of motion gives a value of 3 when 10 times
as many sampling points are used.  However, this improved accuracy comes at the price of computing
time.  Ten times as many sampling points means an increase by a factor of 100 in computing time 
(from about 7 minutes to about 12 hours per orbit).  This made it impractical to use more sampling 
points for each of the 100 orbits.  Despite this scatter, the majority (81\%) of the results lie
between 2.8 and 3.1.  The small cluster of orbits that respect 4 integrals of motion are those
confined to symmetry planes of the potential.
\\
\indent
Another histogram of an ensemble of 100 orbits is shown in Fig. \ref{3dcbhist}.  
These orbits are
supported by the 3D analytical Cazes bar potential and, like orbits in the St\"{a}ckel potential,
there are four groups of 25 orbits at different energies ($\epsilon_J=-0.96$, $-0.85$, $-0.75$, 
$-0.63$).  The initial positions of these orbits have been chosen to cover one quarter of the given
energy surface.  The initial velocities of these orbits have been specified by the Restriction
Hypothesis discussed in \S \ref{verlet}.
\\
\indent
The most conspicuous feature of this histogram is that the integrals of motion are not concentrated 
around a single value.  What this histogram shows is that the 3D Cazes bar potential supports orbits 
that respect 3 integrals of motion (regular), 2 integrals of motion (semiregular), and 1 integral of 
motion (quasi-ergodic).  While this histogram has not been created with a self-consistent stellar
distribution (which is a future project), it does give the flavor of how such a distribution
will appear. 
Additionally, histograms of the 25 orbits at the various Jacobi constants are also interesting.  
They show a trend for an increasing fraction of nonregular orbits with increasing energy.  This is
similar to the trend for orbits in the H\'{e}non-Heiles potential.  
\\
\section{Discussion \& Conclusions}
\indent
Building on the work of \citet{gnp83} and \citet{cns84}, a flexible technique 
for characterizing orbits in 3D potentials known as the correlation integral method has been
presented.  This method analyzes phase space orbits and returns a single
number, the dimensionality of the phase space orbit.  From this number and the 
dimensionality of the underlying phase space, the number of isolating integrals of motion 
respected by an orbit can be determined.  This number can then be used as a quantitative
characterization attribute.
\\
\indent
The implementation of the correlation integral method has been tested for a variety of
previously studied systems.   More familiar characterization tools, such as surfaces of
section and Lyapunov exponents, support the results obtained with the correlation integral
method.  The advantages of the correlation integral are most apparent when used to
characterize orbits in 3D potentials.  A unique, 3D potential based on
a numerically created potential-density pair called the Cazes bar has been presented.  The 
Cazes bar is rotating 
and has no geometrical symmetries that give rise to analytical integrals of motion.  However, 
the correlation integral method 
demonstrates that regular orbits exist in the Cazes bar potential.  Additionally, the
correlation integral method distinguishes between orbits that respect two integrals of
motion and those that respect only one integral.  
\\
\indent  
The simple fact that the correlation integral method can reproduce the results of other 
characterization methods is not enough to warrant its adoption.  Here the various categorizing 
methods discussed in this paper are compared and contrasted.
\begin{itemize}
\item When analyzing orbits in a 4D phase space, surface of section diagrams are the simplest
and clearest way to characterize orbits.  However, there is no simple quantitative measure that
describes quasi-ergodic orbits in surface of section diagrams.  Also, these diagrams are not 
easily translated to 6D phase spaces.  The correlation integral method addresses both of these
problems.
\item Lyapunov exponents provide quantitative measures of orbital regularity
in arbitrary 2D and 3D potentials.  Unfortunately, all regular orbits, closed and unclosed, 
share the same signature in Lyapunov exponent plots.  For orbits in 3D potentials, the behavior
of the Lyapunov exponent is also the same for all nonregular orbits.  That is, no distinction is 
made between orbits that respect only one integral of motion and those that respect two.  The 
correlation integral method distinguishes between these types of regular (periodic and unclosed)
and nonregular (respecting only one or two integrals of motion) orbits.
\item The spectral dynamics characterization method rests on the assumption that the potential
being investigated is at least near-integrable.  When this is true, the spectral dynamics method
provides an excellent way to analyze orbits.  If, instead, the potential is the result of a
numerical simulation or is not clearly near-integrable, then spectral dynamics may not give
relevant results.  Also, spectral dynamics only returns information about regular orbits.  If it
is true that quasi-ergodic orbits play a role in galactic structure, then it will be necessary
to utilize a characterization tool, such as the correlation integral method, that incorporates 
those orbits.
\end{itemize}

The correlation integral method could prove useful for a variety of future studies, but there
are two particular avenues of interest.  Investigations of the impact of massive central objects 
on stellar orbits could benefit from the correlation integral method in the following way.  By
creating histograms, like those shown here in Figs. \ref{stackhist} and \ref{3dcbhist}, from 
orbits in potentials with point masses,
the general characteristics of such a stellar system can be seen.  This information may prove 
useful when creating self-consistent stellar distribution functions to describe black hole/stellar
systems (e.g., van der Marel {\it et al.} 1998, Gebhardt {\it et al.} 2000b).
\\
\indent
The correlation integral method also should 
be a good way to analyze stellar orbits in cosmological simulations that include star formation.
By simply keeping a record of the phase space coordinates of all stars during a simulation, the
correlation integral method can be used to analyze the stellar distribution functions in the
emerging potentials.  To use the Lyapunov exponent method, extra equations would need to be
solved during an already complex calculation.  Similarly, the spectral dynamics technique may
not be expedient if the potentials that arise are not sufficiently close to integrable forms.
\\
\indent
In conclusion, the correlation integral method is a flexible, quantitative, and straightforward 
way of characterizing orbits in 3D potentials.  It is recommended that it be broadly adopted as a tool
for characterizing the properties of orbits in all Hamiltonian dynamical systems.  However, 
there are three specific cases of astrophysical interest to which the correlation
integral method seems particularly well suited:  analyzing stellar distribution functions in
analytically and numerically specified models of steady-state galactic potentials (especially
those with Hamiltonian chaos); investigating the orbital structure supported by galactic potentials
formed in cosmological simulations; and quantifying the response of stellar systems to potentials
that contain central point masses.
\\ 
\acknowledgments
The author would like to thank Joel Tohline, Dana Browne, and Hal Levison for providing
valuable insight on the topics presented here. 
This work has been supported, in part, by funding from the US National Science
Foundation through grant AST 99-87344.  Additional support has been provided by
the Louisiana Board of Regents' LEQSF under agreement LEQSF(1996-01)-GF-08 and
through NASA/LaSPACE under grant NGT5-40035.  This work also has been supported, in
part, by grants of high-performance computing time through NPACI machines at 
the San Diego Supercomputer Center.

\begin{deluxetable}{lcccccccccc} 
\rotate
\tablecolumns{11} 
\tablewidth{0pc} 
\tablecaption{Orbit Information} 
\tablehead{ 
\colhead{Orbit} & \colhead{Fig. \#} & \colhead{$x_0$} & \colhead{$y_0$} & \colhead{$z_0$} 
& \colhead{$\dot{x_0}$} & \colhead{$\dot{y_0}$} & \colhead{$\dot{z_0}$} 
& \colhead{$\epsilon$ or $\epsilon_J$} & \colhead{Lyapunov \tablenotemark{a}} & \colhead{$D$}}
\startdata 
H\'{e}non map & 2 & 0.0 & 0.0 & \nodata & \nodata & \nodata & \nodata & \nodata & \nodata & 1.21\\
2D Richstone \#1 & 3 & 0.5 & 0.0 & \nodata & 0.0 & 0.4 & \nodata & -0.574 & I & 2\\
2D Richstone \#2 & 4 & 0.5 & 0.0 & \nodata & 0.2 & 0.2 & \nodata & -0.614 & I & 2\\
2D Richstone \#3 & 5 & 0.5 & 0.0 & \nodata & 1.12 & 0.23 & \nodata & 0.0 & I & 2\\
2D Richstone \#4 & 6 & 0.136 & -0.532 & \nodata & 0.0 & 0.0 & \nodata & -4.56(-2) & I & 1\\
H\'{e}non-Heiles \#1 & 7 & 0.0 & 0.3 & \nodata & 0.422 & 0.0 & \nodata & 0.125 & I & 1\\
H\'{e}non-Heiles \#2 & 8 & 0.0 & 0.14 & \nodata & 0.481 & 0.0 & \nodata & 0.125 & I & 2\\
H\'{e}non-Heiles \#3 & 9 & 0.0 & 0.24 & \nodata & 0.402 & 0.2 & \nodata & 0.125 & S & 2.4\\
H\'{e}non-Heiles \#4 & 10 & 0.0 & 0.24 & \nodata & 0.281 & 0.35 & \nodata & 0.125 & S & 2.3\\
2D Cazes bar \#1 & 11 & 0.67 & 0.0 & \nodata & 0.0 & 0.611 & \nodata & -0.75 & I & 1\\
2D Cazes bar \#2 & 12 & 0.65 & 0.0 & \nodata & 0.0 & 0.628 & \nodata & -0.75 & I & 2\\
2D Cazes bar \#3 & 13 & 0.5 & 0.0 & \nodata & 0.0 & 0.712 & \nodata & -0.75 & I & 2\\
2D Cazes bar \#4 & 14 & 0.5 & -0.5 & \nodata & 0.0 & 0.507 & \nodata & -0.75 & S & 2.5\\
3D Richstone \#1 & 15 & 0.5 & 0.0 & 0.0 & 0.4 & 0.4 & 0.7 & -0.269 & I & 3\\
3D Richstone \#2 & 16 & 0.6 & 0.0 & 0.2 & 0.5 & 0.2 & -0.05 & -0.263 & I & 3\\
3D Richstone \#3 & 17 & 0.5 & 0.0 & 0.3 & 0.4 & 0.5 & 0.01 & -0.229 & I & 3\\
\enddata 
\tablenotetext{a}{The letters that appear in this column distinguish between orbits that are
insensitive (I) to small changes in initial conditions and those that are sensitive (S) to
such changes.  All orbits labeled `I' are regular.}
\end{deluxetable}

\begin{figure} 
\caption{(a) Equipotential contours of the numerical Cazes bar in the $x-z$ 
plane for $x>0$; (b) equipotential contours of the analytical Cazes bar in the same 
plane.  (c) Equipotential contours of the numerical Cazes bar in the $y-z$ 
plane for $y>0$; (d) equipotential contours of the analytical Cazes bar in the same 
plane.  (e) Equipotential contours of the numerical Cazes bar in the $x-y$ 
plane; (f) equipotential contours of the analytical Cazes bar in the same plane.
\label{acbvsncb}}
\end{figure}

\begin{figure}
\caption{H\'{e}non map.  (a) Plot of phase space points obtained after 100,000 iterations 
of eqs. (\ref{henx}) and (\ref{heny}).  (b) Magnification of the plot in (a).  The small 
scale substructure is a telltale sign of the fractal nature of the phase space orbit.  
(c) A plot of the correlation integral ($\ln C(r)$) versus phase space distance ($\ln r$) 
for this mapping.  The value of $D$ shown in the legend has been measured from the slope of 
the line; $\sigma$ is the standard deviation measured for $D$, as described in \S 
\ref{introcorr}.\label{henplot}}
\end{figure}

\begin{figure}
\caption{Orbit \#1 in the 2D Richstone potential; see Table 1 for initial conditions.  (a) 
The $R-z$ trajectory of this orbit.  (b) The
associated $C-r$ plot listing the average slope ($D$) and standard deviation ($\sigma$).
(c) The plot of the Lyapunov exponent ($\ln k_n$) versus integration time ($\ln n\tau$) for
this orbit.  (d) The ($R-p_R$) surface of section for this orbit.
\label{2drich1}} 
\end{figure}

\begin{figure}
\caption{Orbit \#2 in the 2D Richstone potential; each frame contains information as described
in the caption to Fig. \ref{2drich1}.\label{2drich2}} 
\end{figure}

\begin{figure}
\caption{Orbit \#3 in the 2D Richstone potential; each frame contains information as described
in the caption to Fig. \ref{2drich1}.\label{2drich3}} 
\end{figure}

\begin{figure}
\caption{Orbit \#4 in the 2D Richstone potential; each frame contains information as described
in the caption to Fig. \ref{2drich1}.\label{2drich4}} 
\end{figure}

\newpage

\begin{figure}
\caption{Orbit \#1 in the H\'{e}non-Heiles potential; see Table 1 for initial conditions.  (a) 
The $R-z$ trajectory of this orbit.  (b) The
associated $C-r$ plot listing the average slope ($D$) and standard deviation ($\sigma$).
(c) The plot of the Lyapunov exponent ($\ln k_n$) versus integration time ($\ln n\tau$) for
this orbit.  (d) The ($R-p_R$) surface of section for this orbit.
\label{henh1}} 
\end{figure}

\begin{figure}
\caption{Orbit \#2 in the H\'{e}non-Heiles potential; each frame contains information as described
in the caption to Fig. \ref{henh1}.\label{henh2}} 
\end{figure}

\begin{figure}
\caption{Orbit \#3 in the H\'{e}non-Heiles potential; each frame contains information as described
in the caption to Fig. \ref{henh1}.\label{henh3}} 
\end{figure}

\begin{figure}
\caption{Orbit \#4 in the H\'{e}non-Heiles potential; each frame contains information as described
in the caption to Fig. \ref{henh1}.\label{henh4}} 
\end{figure}

\begin{figure}
\caption{Orbit \#1 in the 2D Cazes bar potential; see Table 1 for initial conditions.  (a) 
The $x-y$ trajectory of this orbit.  (b) The 
associated $C-r$ plot listing the average slope ($D$) and standard deviation ($\sigma$).
(c) The plot of the Lyapunov exponent ($\ln k_n$) versus integration time ($\ln n\tau$) for
this orbit.  (d) The ($x-p_x$) surface of section for this orbit.
\label{2dcazes1}}
\end{figure}

\begin{figure}
\caption{Orbit \#2 in the 2D Cazes bar potential; each frame contains information as described
in the caption to Fig. \ref{2dcazes1}.\label{2dcazes2}} 
\end{figure}

\newpage

\begin{figure}
\caption{Orbit \#3 in the 2D Cazes bar potential; each frame contains information as described
in the caption to Fig. \ref{2dcazes1}.\label{2dcazes3}} 
\end{figure}

\begin{figure}
\caption{Orbit \#4 in the 2D Cazes bar potential; each frame contains information as described
in the caption to Fig. \ref{2dcazes1}.\label{2dcazes4}} 
\end{figure}

\begin{figure}
\caption{Orbit \#1 in the 3D Richstone potential; see Table 1 for initial conditions.  (a) 
The $x-y$ plane projection of this orbit.  
(b) The $x-z$ plane projection of the same orbit.  (c) The $y-z$ plane projection of the 
same orbit.  (d) The associated $C-r$ plot listing the average slope ($D$) and standard
deviation ($\sigma$). (e) The plot of the Lyapunov exponent ($\ln k_n$) versus integration
time ($\ln n\tau$) for this orbit.
\label{3drich1}}
\end{figure}

\begin{figure}
\caption{Orbit \#2 in the 3D Richstone potential; each frame contains information as described
in the caption to Fig. \ref{3drich1}.\label{3drich2}} 
\end{figure}

\begin{figure}
\caption{Orbit \#3 in the 3D Richstone potential; each frame contains information as described
in the caption to Fig. \ref{3drich1}.\label{3drich3}} 
\end{figure}

\begin{figure}
\caption{Histogram of the number of orbits versus the number of integrals of motion for 100 orbits 
in the St\"{a}ckel potential described in \S \ref{stack}.\label{stackhist}}
\end{figure}

\newpage

\begin{figure}
\caption{Histogram of the number of orbits versus the number of integrals of motion for 100 orbits
in the 3D analytical Cazes bar potential.\label{3dcbhist}}
\end{figure}

\end{document}